\documentclass{article}
\usepackage{amssymb}
\usepackage{amsmath}

\begin{document}

\title{\large In-Out Formalism for One-Loop Effective Actions in QED and Gravity}
\author{Sang Pyo Kim\thanks{%
Department of Physics, Kunsan National University, Kunsan 54150, Korea; e-mail: sangkim@kunsan.ac.kr}}
\maketitle

\begin{abstract}
The in-out formalism is a systematic and powerful method for finding the effective actions in an electromagnetic field and a curved spacetime provided that the field equation has explicitly known solutions. The effective action becomes complex when pairs of charged particles are produced due to an electric field and curved spacetime. This may lead to a conjecture of one-to-one correspondence between the vacuum polarization (real part) and the vacuum persistence (imaginary part). We illustrate the one-loop effective action in a constant electric field in a Minkowski spacetime and in a uniform electric field in a two-dimensional (anti-) de sitter space.
\end{abstract}

\section{Introduction} \label{sec-int}

The spontaneous production of charged pairs in an electromagnetic field or in a curved space time has been one of the key issue in theoretical physics since the pioneering work by Heisenberg-Euler \cite{heisenberg-euler}, Schwinger \cite{schwinger}, Parker \cite{parker}, and Hawking \cite{hawking}. The vacuum polarization (one-loop effective action) has been found only for a constant electromganetic field \cite{heisenberg-euler,schwinger} or Sauter-type electric or magnetic field \cite{kim-lee-yoon08,kim11}. The vacuum polarization in a Schwarzchild black hole has recently been found in the Schwinger proper-time integral \cite{kim-hwang}. Quantum effects in strong external fields have been intensively studied in Ref.~\cite{grib-mamaev-mostepanenko} and
also the propagator approach has been applied to QED in Ref.~\cite{fradkin-gitman-shvartsman}.

In this article to the proceedings we elaborate further the in-out formalism to find the one-loop effective action in a curved spacetime as well as an electromagnetic field, which respects the spacetime symmetry, in particular, a uniform electric field in an (anti-) de Sitter (dS) space. In the in-out formalism advanced by Schwinger and DeWitt \cite{dewitt}, the scattering matrix of the out-vacuum with respect to the out-vacuum gives the one-loop effective action
\begin{eqnarray}
e^{i {\cal W}} = e^{i \int \sqrt{-g} d^D x {\cal L}_{\rm eff}} = \langle 0, {\rm out} \vert 0, {\rm in} \rangle,
\end{eqnarray}
which is equivalent to the Feynman integral or infinite sum of all Feynman diagrams with even number of external photons and/or gravitons. In particular, when the vacuum becomes unstable due to the production of particles or pairs, the effective action becomes complex and leads to the consistent relation
\begin{eqnarray}
\vert \langle 0, {\rm out} \vert 0, {\rm in} \rangle \vert^2 = e^{- 2\, {\rm Im} ({\cal W})}, \quad 2\, {\rm Im} \bigl( {\cal W} \bigr) = \pm \bigl( {\cal VT} \bigr) \sum_{\kappa} \ln \bigl(1 \pm {\cal N}_{\kappa} \bigr)
\end{eqnarray}
where ${\cal VT}$ is the spacetime valume and ${\cal N}_{\kappa}$ is the mean number of produced pairs carrying quantum number ${\kappa}$. Thus, the pair production is closely and consistently related to the vacuum polarization.

The scattering matrix can be found from the Bogoliubov transformation between the in-vacuum and the out-vacuum
\begin{eqnarray}
\hat{a}_{\kappa , {\rm out}} = \alpha_{\kappa} \hat{a}_{\kappa , {\rm in}} + \beta_{\kappa} \hat{b}^{\dagger}_{\kappa , {\rm in}}, \quad
\hat{b}_{\kappa , {\rm out}} = \alpha_{\kappa} \hat{b}_{\kappa , {\rm in}} + \beta_{\kappa} \hat{a}^{\dagger}_{\kappa , {\rm in}},
\end{eqnarray}
where $\hat{a}_{\kappa , {\rm in}}$ and $\hat{b}_{\kappa , {\rm in}}$ are the particle and anti-particle operators in the in-vacuum and $\hat{a}_{\kappa , {\rm out}}$ and $\hat{b}_{\kappa , {\rm out}}$ are those in the out-vacuum.
Then, the out-vacuum for bosons is superposed of the multi-particle and anti-particle states of the in-vacuum
\begin{eqnarray}
\vert 0, {\rm in} \rangle = \prod_{\kappa} \frac{1}{\alpha_{\kappa}} \sum_{n_{\kappa}} \Bigl( - \frac{\beta^*_{\kappa}}{\alpha_{\kappa}} \Bigr)^{n_{\kappa}} \vert n_{\kappa}, \bar{n}_{\kappa}, {\rm in} \rangle,
\end{eqnarray}
while due to the Pauli blocking the out-vacuum for fermions takes the form
\begin{eqnarray}
\vert 0, {\rm in} \rangle = \prod_{\kappa} \frac{1}{\alpha_{\kappa}} \Bigl( - \beta^*_{\kappa} \vert 1_{\kappa}, \bar{1}_{\kappa}, {\rm in} \rangle + \alpha_{\kappa} \vert 0_{\kappa}, \bar{0}_{\kappa}, {\rm in} \rangle \Bigr).
\end{eqnarray}
Therefore, the integrated one-loop effective action is given by
\begin{eqnarray} \label{sc mat}
{\cal W} = \pm i \bigl( {\cal VT} \bigr) \sum_{\kappa} \ln \bigl( \alpha^*_{\kappa} \bigr),
\end{eqnarray}
where the upper (lower) sign is for bosons (fermions). The explicit form of $\alpha_{\kappa}$ gives rise to, in principle, the one-loop action.

\section{Reconstructing QED Action} \label{sec-recon}

In quantum field theory, the correspondence has been known between the imaginary part and the real part of physical quantities, such as the dispersion relation, which is a consequence of the analytical nature of the underlying quantity in a complex plane. One may thus raise a question whether one can find the effective action from the pair production, the former corresponding to the real part and the latter to the imaginary part of the effective action. The derivation of the imaginary part from the real part of the effective action is rather straightforward in the proper-time integral representation of the one-loop action since the simple poles determine the imaginary part. This corresponds to the inverse procedure of the Borel summation \cite{dunne-hall,dunne-schubert}, which cannot, however, always apply to the models considered in this paper. Analyzing many quantum field models that have been exactly solved, the author and Schubert have observed that the imaginary part (vacuum persistence) of the effective action may be factorized into a product of one plus or minus exponential factors
\begin{eqnarray}
2\, {\rm Im} \bigl( {\cal L}^{(1)}_{\rm eff} \bigr) = \sum_{\kappa} \sum_{I_{\kappa}} \pm \ln \bigl(1\pm e^{- S^{I_{\kappa}}_{\kappa}} \bigr), \label{per con}
\end{eqnarray}
where $\kappa$ and $I_{\kappa}$ denote quantum numbers and the kinds of instanton actions for the field. Then, the structure of simple poles and their residues leading to these factors uniquely determine the analytical structure of the proper-time integrand of the effective action (vacuum polarization) modulo some entire functions via the Mittag-Leffler theorem. The entire functions are independent of renormalization of physical quantities, so they are to be regulated away.

Based on the above observation, the author and Schubert have proposed the following correspondence between the imaginary part and the real part of the effective action \cite{kim-schubert}
\begin{eqnarray} \label{cor}
\pm i \ln ( 1 \pm e^{- S}) \Leftrightarrow P \int_{0}^{\infty} \frac{ds}{s} e^{-\frac{Ss}{\pi}} \Bigl( \frac{(\cos s)^{2 |\sigma|}}{\sin s} - f_{\sigma} (s) \Bigr),
\end{eqnarray}
where $P$ and $\sigma$ denote the principal value and the spin $(\sigma = 0, 1/2)$, and the regulating function $f_{\sigma} (s) = 1/s - \cdots$ is subtracted ala the Schwinger scheme and corresponds to the renormalization of physical quantities. The explicit form of $f_{\sigma}(s)$  is determined after the proper sum of all quantum numbers, such as the momentum and spin etc, so that the renormalized one-loop effective action should be finite in the proper-time integral. The upper (lower) sign corresponds to the scalar (spinor) QED action in a constant electric field. Note that the vacuum persistence for the Bose-Einstein and Fermi-Dirac distributions
\begin{eqnarray}
2\, {\rm Im} \bigl({\cal W}_{\kappa} \bigr) = \pm \ln \Bigl(1 \pm \frac{1}{e^{S_{\kappa}} \mp 1} \Bigr) = \mp \ln \bigl(1 \mp e^{-S_{\kappa}} \bigr),
\end{eqnarray}
leads to the opposite spin-statistics as afar as the one-loop effective action is concerned. This fact has been first pointed out by Stephens \cite{stephens} and then discussed in detail in Ref.~\cite{hwang-kim}. The effective action density may be found by summing over proper density of states.

We illustrate the conjecture with the Schwinger effect and the Heisenberg-Euler and Schwinger action. The mean number for produced pairs with charge $q$ and the transverse momentum ${\bf k}_{\perp}$ in a constant electric field $E$ in a $D$-dimensional spacetime is given by
\begin{eqnarray}
{\cal N}_{{\bf k}_{\perp}} = e^{- \pi \frac{m^2 + {\bf k}_{\perp}^2}{qE}}
\end{eqnarray}
regardless of spins. Using the density of states
\begin{eqnarray}
{\cal D}_{{\bf k}_{\perp} \sigma} = \frac{(2 \vert \sigma \vert +1)}{2} \Bigl( \frac{qE}{2\pi} \Bigr) \frac{d^{D-2}{\bf k}_{\perp}}{(2 \pi)^{D-2}},
\end{eqnarray}
the effective action follows from Eq.~(\ref{cor}) as
\begin{eqnarray}
{\cal L}^{(1)}_{\rm eff} = \frac{(2 \vert \sigma \vert +1)}{2} \Bigl( \frac{qE}{2\pi} \Bigr) \int \frac{d^{D-2}{\bf k}_{\perp}}{(2 \pi)^{D-2}} \int_{0}^{\infty} \frac{ds}{s} e^{-\frac{m^2 + {\bf k}_{\perp}^2}{qE} s} \Bigl( \frac{(\cos s)^{2 |\sigma|}}{\sin s} - f_{\sigma} (s) \Bigr).
\end{eqnarray}
When $D=4$, for instance, the momentum integral leads to $qE/(2 \pi s^2)$ and the regulating function should be $f_0 (s) = 1/s + s/3$ and $f_{1/2} (s) = 1/s - s/3$, in which each term renormalizes the vacuum energy and charge.

The Sauter-type pulsed or localized electric field provides an intriguing model for reconstructing the one-loop QED action. The mean number in $E(t) = E_0 / \cosh^2 (t/T)$ has been known and leads to the vacuum persistence amplitude \cite{kim-lee-yoon08} (and references therein)
\begin{eqnarray}
2\, {\rm Im} \bigl( {\cal W}_{\bf k} \bigr) = (-1)^{2 |\sigma|} \bigl(1 + 2 |\sigma| \bigr) \ln \Bigl(\frac{\bigl(1 + (-1)^{2|\sigma|} e^{- \pi \Omega_{(+)}} \bigr)\bigl(1 + (-1)^{2|\sigma|} e^{- \pi \Omega_{(-)}} \bigr)}{\bigl(1 - e^{- 2 \pi T \omega_{(+)}} \bigr)\bigl(1 - e^{- 2 \pi T \omega_{(-)}} \bigr)} \Bigr),
\end{eqnarray}
where
\begin{eqnarray}
\omega_{(\pm)} &=& \sqrt{\bigl(k_{\parallel} \mp qE_0 T \bigr)^2 + m^2 + {\bf k}_{\perp}^2}, \nonumber\\
\Omega_{(\pm)} &=& T \bigl( \omega_{(+)} + \omega_{(-)}  \bigr) \pm 2 \bigl(qE_0 T^2 \bigr) \sqrt{1 - \frac{1 - 2 |\sigma|}{\bigl(qE_0 T^2 \bigr)^2}}.
\end{eqnarray}
On the other hand, the mean number in $E(x) = E_0 / \cosh^2 (x/L)$ leads to the vacuum persistence amplitude \cite{kim-lee-yoon08}
\begin{eqnarray}
2\, {\rm Im} \bigl( {\cal W}_{\omega {\bf k}_{\perp}} \bigr) = (-1)^{2 |\sigma|} \bigl(1 + 2 |\sigma| \bigr) \ln \Bigl(\frac{\bigl(1 + (-1)^{2|\sigma|} e^{- \pi \Omega_{(+)}} \bigr)\bigl(1 + (-1)^{2|\sigma|} e^{\pi \Omega_{(-)}} \bigr)}{\bigl(1 + (-1)^{2|\sigma|} e^{- \pi \Delta_{(+)}} \bigr)\bigl(1 +(-1)^{2|\sigma|}  e^{\pi \Delta_{(-)}} \bigr)} \Bigr),
\end{eqnarray}
where
\begin{eqnarray}
p_{\parallel (\pm)} &=& \sqrt{\bigl(\omega_{\parallel} \mp qE_0 L \bigr)^2 - m^2 - {\bf k}_{\perp}^2}, \nonumber\\
\Omega_{(\pm)} &=& L \bigl( p_{\parallel (+)} + p_{\parallel (-)}  \bigr) \pm 2 \bigl(qE_0 L^2 \bigr) \sqrt{1 - \frac{1 - 2 |\sigma|}{\bigl(qE_0 L^2 \bigr)^2}}, \nonumber\\
\Delta_{(\pm)} &=& L \bigl( p_{\parallel (+)} - p_{\parallel (-)}  \bigr) \pm 2 \bigl(qE_0 L^2 \bigr) \sqrt{1 - \frac{1 - 2 |\sigma|}{\bigl(qE_0 L^2 \bigr)^2}}.
\end{eqnarray}
Hence, the vacuum polarization from the correspondence (\ref{cor}) coincides with that directly from Eq.~(\ref{sc mat}) in the in-out formalism \cite{kim-lee-yoon08}.

\section{QED in (A)dS${}_2$} \label{sec-qed in ads}

Next, we illustrate the QED action in a uniform electric field in ${\rm (A)dS}_2$. There are two coordinates covering ${\rm (A)dS}_2$. First, in the global coordinates, the metrics are given, respectively, by
\begin{eqnarray}
ds^2_{\rm dS} = - dt^2 + \cosh^2 (Ht) dx^2, \quad ds^2_{\rm AdS} = - \cosh^2 (Kx) dt^2 + dx^2.
\end{eqnarray}
The mean number for the Schwinger effect \cite{kim-page08} in the global coordinates has been recalculated in Ref.~\cite{kim-hwang-wang}
\begin{eqnarray}
{\cal N}_{\rm dS} (E) = \Bigl(\frac{\cosh Y}{\sinh X} \Bigr)^2, \quad {\cal N}_{\rm AdS} (E) = \Bigl(\frac{\sinh \tilde{X}}{\cosh \tilde{Y}} \Bigr)^2,
\end{eqnarray}
where
\begin{eqnarray}
X &=& 2 \pi \sqrt{\Bigl(\frac{qE}{R_{\rm dS}} \Bigr)^2+ \frac{m^2}{2 R_{\rm dS}} - \frac{1}{16}}, \quad Y = 2 \pi \frac{qE}{R_{\rm dS}}, \nonumber\\
\tilde{X} &=& 2 \pi \sqrt{\Bigl(\frac{qE}{R_{\rm AdS}} \Bigr)^2+ \frac{m^2}{2 R_{\rm AdS}} - \frac{1}{16}}, \quad \tilde{Y} = - 2 \pi \frac{qE}{R_{\rm AdS}}.
\end{eqnarray}
The scalar curvature is $R_{\rm dS}= 2H^2$ and $R_{\rm AdS} = - 2K^2$ and $S_{\rm dS} = 2X$ and $S_{\rm AdS} = 2\, \tilde{X}$ are instanton actions from the field equation or phase-integral method. Noting the density of states ${\cal D}_{\rm dS} = (2 H^2X)/(2 \pi)^2$, we find the QED action in ${\rm dS}_2$ from the reconstruction conjecture in Sec.~\ref{sec-recon}
\begin{eqnarray}
{\cal L}_{\rm dS}^{(1)} (E) &=& \frac{H^2 X}{2 \pi^2} \Biggl[P \int_{0}^{\infty} \frac{ds}{s} e^{-\frac{2Xs}{\pi}} \cosh \bigl( \frac{2Ys}{\pi} \bigr) \Bigl( \frac{1}{\sin s} - \frac{1}{s} \Bigr)\nonumber\\
&& - P \int_{0}^{\infty} \frac{ds}{s} e^{-\frac{2Xs}{\pi}} \Bigl( \frac{\cos s}{\sin s} - \frac{1}{s} \Bigr) \Biggr],
\end{eqnarray}
Similarly, using the density of states ${\cal D}_{\rm AdS} = (2 K^2X)/(2 \pi)^2$, we find the QED action in ${\rm AdS}_2$
\begin{eqnarray}
{\cal L}_{\rm AdS}^{(1)} (E) = \frac{K^2 \tilde{X}}{2 \pi^2} P \int_{0}^{\infty} \frac{ds}{s} e^{-\frac{2\tilde{Y}s}{\pi}} \Bigl(\cosh \bigl( \frac{2\tilde{X}s}{\pi} \bigr) -  1 \Bigr) \Bigl( \frac{1}{\sin s} - \frac{1}{s} \Bigr).
\end{eqnarray}

Second, we consider the planar coordinates for ${\rm (A)dS}_2$ with the metrics
\begin{eqnarray}
ds^2_{\rm dS} = - dt^2 + e^{2Ht} dx^2, \quad ds^2_{\rm AdS} = - e^{2Kx} dt^2 + dx^2.
\end{eqnarray}
The mean number for ${\rm dS}_2$ is given by \cite{cai-kim14}
\begin{eqnarray}
{\cal N}_{\rm dS} = \frac{e^{- 2 X + 2Y}+ e^{-4 X} }{1 - e^{-4X}},
\end{eqnarray}
while the mean number for ${\rm AdS}_2$ is given by
\begin{eqnarray}
{\cal N}_{\rm AdS} = \frac{e^{- 2 \tilde{Y} + 2 \tilde{X}} - e^{- 2\tilde{Y} - 2 \tilde{X}} }{1 + e^{-2\tilde{Y} - 2 \tilde{X}}}.
\end{eqnarray}
From the reconstruction conjecture, we thus obtain the one-loop effective action for ${\rm dS}_2$
\begin{eqnarray} \label{qed ds pl}
{\cal L}_{\rm dS}^{(1)} &=& \frac{H^2 X }{(2 \pi)^2} \Biggl[P \int_{0}^{\infty} \frac{ds}{s} e^{-\frac{2(X-Y)s}{\pi}} \Bigl( \frac{1}{\sin s} - \frac{1}{s} \Bigr) \nonumber\\
&& \qquad - P \int_{0}^{\infty} \frac{ds}{s} e^{-\frac{4Xs}{\pi}} \Bigl( \frac{\cos s}{\sin s} - \frac{1}{s} \Bigr) \Biggr],
\end{eqnarray}
and for ${\rm AdS}_2$
\begin{eqnarray} \label{qed ads pl}
{\cal L}_{\rm AdS}^{(1)} &=& \frac{K^2 \tilde{X}}{(2 \pi)^2} P \int_{0}^{\infty} \frac{ds}{s} e^{-\frac{2\tilde{Y}s}{\pi}} \cosh \Bigl(
\frac{2\tilde{X} s}{\pi} \Bigr) \Bigl( \frac{1}{\sin s} - \frac{1}{s} \Bigr).
\end{eqnarray}
The QED actions in Eqs.~(\ref{qed ds pl}) and (\ref{qed ads pl}) are the same as those in Ref.~\cite{cai-kim14}, which have been derived from the scattering matrix (\ref{sc mat}) via the gamma-function regularization \cite{kim-lee-yoon08}.

The QED actions have the duality of ${\rm R}_{\rm dS} \rightleftarrows {\rm R}_{\rm AdS}$, while the effective actions in pure ${\rm dS}_2$ and ${\rm AdS}_2$ do not have the duality. In the limit of zero electric field $(E=0)$, one obtains the mean number in the global coordinates of ${\rm dS}_2$
\begin{eqnarray}
{\cal N}_{\rm dS} = \frac{1}{\bigl(\sinh X \bigr)^2}
\end{eqnarray}
and in the planar coordinates
\begin{eqnarray}
{\cal N}_{\rm dS} = \frac{1}{e^{2X} -1}.
\end{eqnarray}
Here $X = \pi \sqrt{(m/H)^2 - 1/4}$. We thus obtain the effective action for pure ${\rm dS}_2$ in the global coordinates
\begin{eqnarray} \label{pure ds gl}
{\cal L}_{\rm dS}^{(1)} = \frac{H^2 X}{2 \pi^2} P \int_{0}^{\infty} \frac{ds}{s} e^{-\frac{2Xs}{\pi}} \frac{\sin(s/2)}{\cos (s/2)},
\end{eqnarray}
and in the planar coordinates
\begin{eqnarray} \label{pure ds pl}
{\cal L}_{\rm dS}^{(1)} = \frac{H^2 X}{(2 \pi)^2} P \int_{0}^{\infty} \frac{ds}{s} e^{-\frac{2Xs}{\pi}} \Bigl( \frac{\cos (s)}{\sin (s)} - \frac{1}{s} \Bigr).
\end{eqnarray}
Note that the spectral function in Eq.~(\ref{pure ds gl}) is neither the spinor nor scalar QED action form while the effective action in the planar coordinates is that for spinor QED. The one-loop action may be compared with the action in all dimensions in Ref.~\cite{das-dunne}.

\section{Conclusions} \label{sec-con}
We have elaborated the in-out formalism for finding the one-loop effective action in an electric field and curved spacetime. The one-loop action takes a complex value when the vacuum becomes unstable against pair production. The vacuum persistence, the twice of the imaginary part, is determined by the mean number of produced pairs and may lead to the vacuum polarization, the real part of the effective action through the reconstruction conjecture.

We have illustrated the QED actions from the mean number in a constant electric field and Sauter-type pulsed or localized electric field. We have then obtained the QED actions in a uniform electric field in the global coordinates and planar coordinates for a two-dimensional de Sitter or anti-de Sitter space. The QED actions from the correspondence recover those directly from the scattering matrix in the in-out formalism. Finally, the one-loop actions have been found for pure de Sitter space in the both coordinates.

\section*{Acknowledgements}
The author would like to thank Rong-Gen Cai, W-Y.~Pauchy Hwang, Hyun Kyu Lee, Christian Schubert and Yongsung Yoon for collaborations.
This work was supported by Basic Science Research Program through the National Research Foundation of Korea (NRF) funded by the Ministry of Education (15B15770630).

\end{document}